\documentclass[a4paper]{jpconf}
\usepackage{graphicx}
\usepackage{amssymb}
\usepackage{xspace}

\begin{document}

\title{Hadron yields, the chemical freeze-out and the QCD phase diagram}

\author{A Andronic$^1$, P Braun-Munzinger$^{1,2}$, K Redlich$^{3}$, and J Stachel$^2$}

\address{$^1$~Research Division and EMMI, GSI Helmholtzzentrum f\"ur Schwerionenforschung, Darmstadt, Germany}
\address{$^2$~Physikalisches Institut, Universit\"at Heidelberg, Germany}
\address{$^3$~Institute of Theoretical Physics, University of Wroc\l aw, Poland}

\ead{A.Andronic@gsi.de}

\begin{abstract}
We present the status of the chemical freeze-out, determined from fits of hadron yields with the statistical hadronization (thermal) model, with focus on the data at the LHC. A description of the yields of hadrons containing light quarks as well as the application of the model for the production of the J/$\psi$ meson is presented. The implications for the QCD phase diagram are discussed.

\end{abstract}

\section{Chemical freeze-out of light-quark hadrons}

Chemical freeze-out in nucleus-nucleus collisions is addressed in the statistical hadronization (thermal) model, based on the statistical operator for the hadron resonance gas. It describes a snapshot of the collision dynamics, namely assuming a rapid chemical freeze-out.
The model is simple, but very powerful considering its small set of parameters
(temperature $T$, baryochemical potential $\mu_B$ and volume $V$). 
Importantly, the model is a unique phenomenological approach linking
the production of hadrons in nucleus-nucleus collisions to the QCD
phase diagram \cite{BraunMunzinger:2008tz} (see
\cite{Cabibbo:1975ig,Hagedorn:1984hz} for early ideas).

All known hadrons, stable and decaying, are employed in the
calculations.  Even as the knowledge of the hadron spectrum is
constantly improving, missing states could in principle still have an
effect on $T$, as shown in lattice QCD studies
\cite{Bazavov:2014xya}. 
As currently the prefered $T$ values are around 160 MeV (see below)
the relevance of higher-lying resonances is diminished.  In the
commonly-used grand canonical approach, chemical potentials $\mu$
ensure conservation on average of additive quantum numbers (baryon
number, isospin, strangeness, charmness), fixed by ``initial
conditions".

The model, in its ``standard" implementation (meaning the minimal set
of parameters listed above), was sucessfully used to describe hadron
production in heavy-ion collisions over a wide range of collision
energies (see e.g. \cite{Andronic:2005yp,Becattini:2005xt}).
Several versions with extended set of parameters have been proposed
\cite{Letessier:2005qe,Petran:2013qla}. 
In particular, a strangeness suppression factor $\gamma_s$ is used as
a fit parameter to test the possible departure from equilibrium of hadrons
containing strange quark(s) (see, e.g. \cite{Becattini:2005xt}).
The question of flavor-dependent freeze-out is also addressed
\cite{Bazavov:2013dta,Bellwied:2013cta,Chatterjee:2013yga}.  Beyond
the sudden freeze-out concept, a hadronic phase with ``chemical
activity'' with the UrQMD transport model \cite{Becattini:2016xct} was
proposed.  Some times, possible repulsive interactions among hadrons
are modeled in a hard-sphere excluded-volume approach.  In the
simplest implementation, identical (and moderate) values for the radii
are used ($R_{meson}=R_{baryon}=0.3$ fm), leaving $T$ and $\mu_B$
unaffected in comparison with results for point-like hadrons
\cite{Andronic:2005yp}. Employing species-dependent hard-sphere
interactions \cite{Vovchenko:2015cbk} leads to an extended parameter
set, which, however, cannot be constrained by current knowledge of
hadron-hadron interactions. 
As the results below demonstrate, the data do not require any of these extensions.

For small systems and/or low energies, a canonical treatment is needed
\cite{Cleymans:1998yb}, usually implemented only for strangeness.  
Recent such studies in p--Nb and Ar--KCl collisions
\cite{Agakishiev:2015bwu} and in pp collisions
\cite{Vovchenko:2015idt,Das:2016muc,Cleymans:2016qnc} lead to values
of $T$ comparable to (or even larger than) those in (central) Au--Au
or Pb--Pb collisions. The studies 
performed by ALICE in p--Pb \cite{Adam:2015vsf} and pp
\cite{Adam:2016emw} collisions revealed that in high-multiplicity
events hadron production in these small systems resembles that in
Pb--Pb collisions.

\begin{figure}[htb]
\begin{minipage}{0.7\textwidth}
\centerline{\hspace{-.5cm}\includegraphics[width=.99\textwidth]{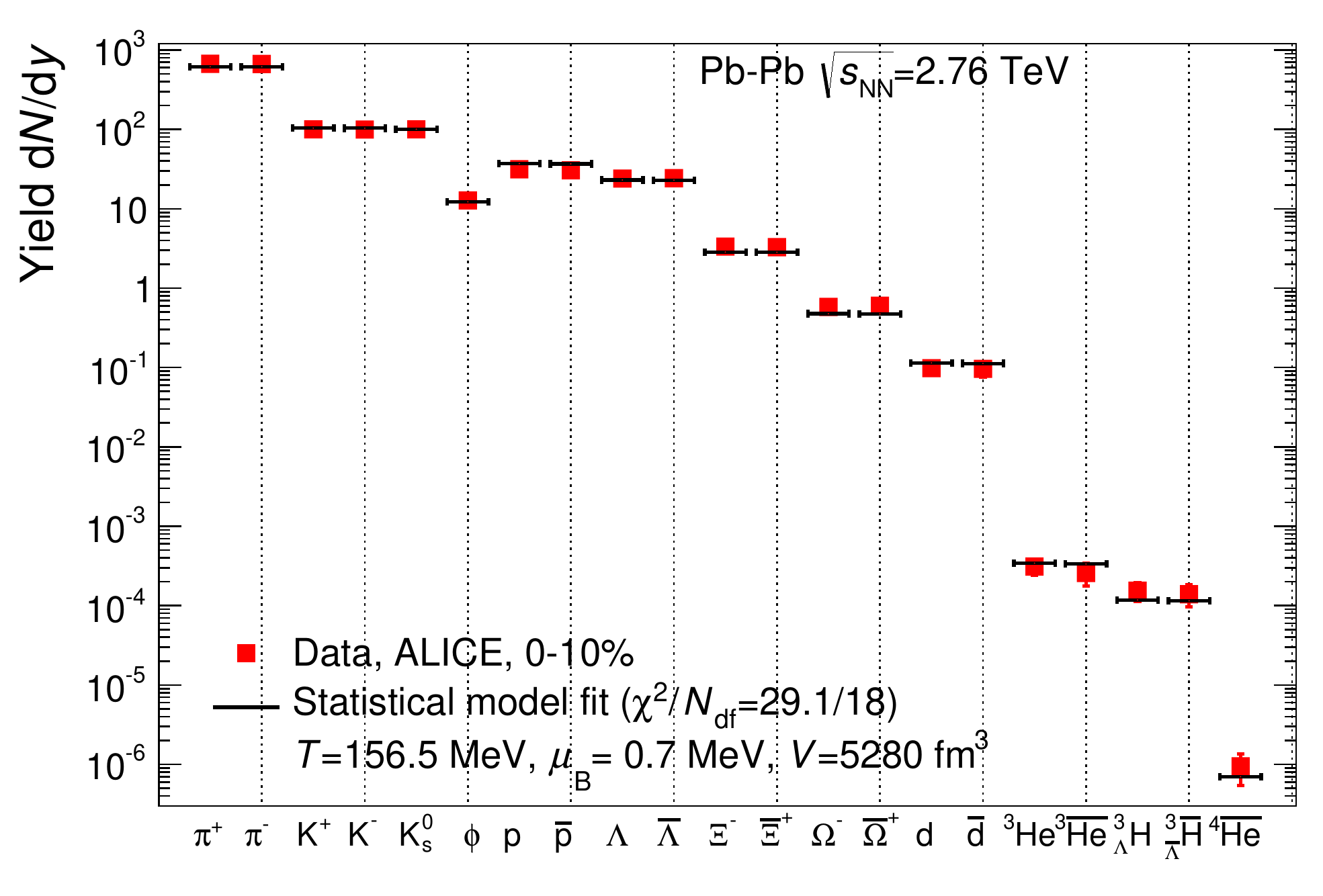}}

\centerline{\mbox{~~~~~}\includegraphics[width=.96\textwidth]{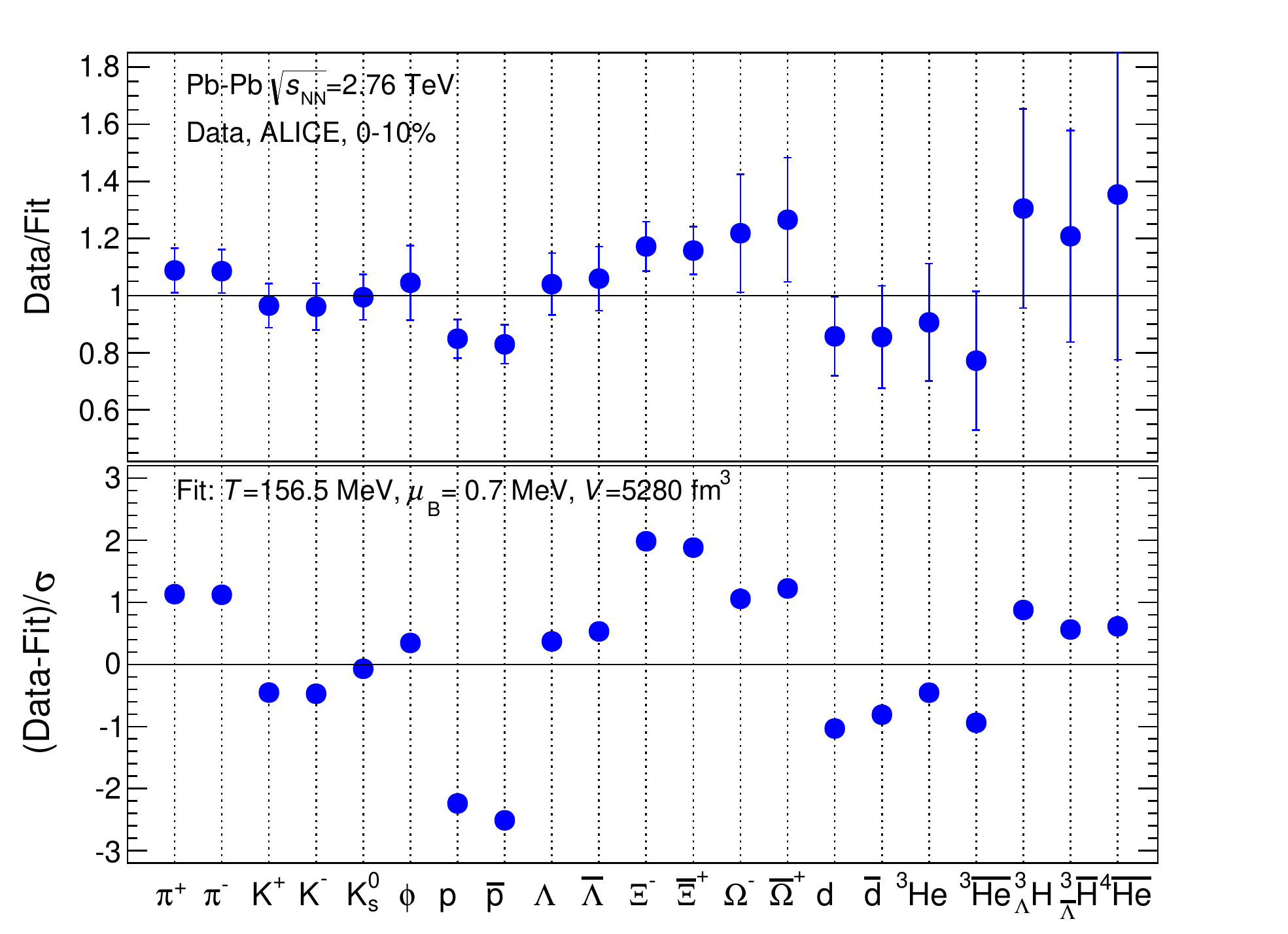}}
\end{minipage} \begin{minipage}{0.3\textwidth}

\caption{\label{f:fit}Hadron multiplicities in central (0-10\%) Pb--Pb collisions at the LHC, ALICE data (\cite{Abelev:2012wca,Abelev:2013vea,Abelev:2013xaa,ABELEV:2013zaa,Abelev:2014uua,Adam:2015yta,Adam:2015vda}; the values for $\bar{\Lambda}$ \cite{Schuchmann:2102194} and anti-$^4$He are preliminary) and best fit (upper panel). The lower panels show the ratio of data value to fit (with error bars the total, statistical and systematic, uncertainties) and the difference between data and model fit in units of the experimental uncertainty.}
\end{minipage} 
\end{figure}

\begin{figure}[htb]
\centerline{\includegraphics[width=.7\textwidth]{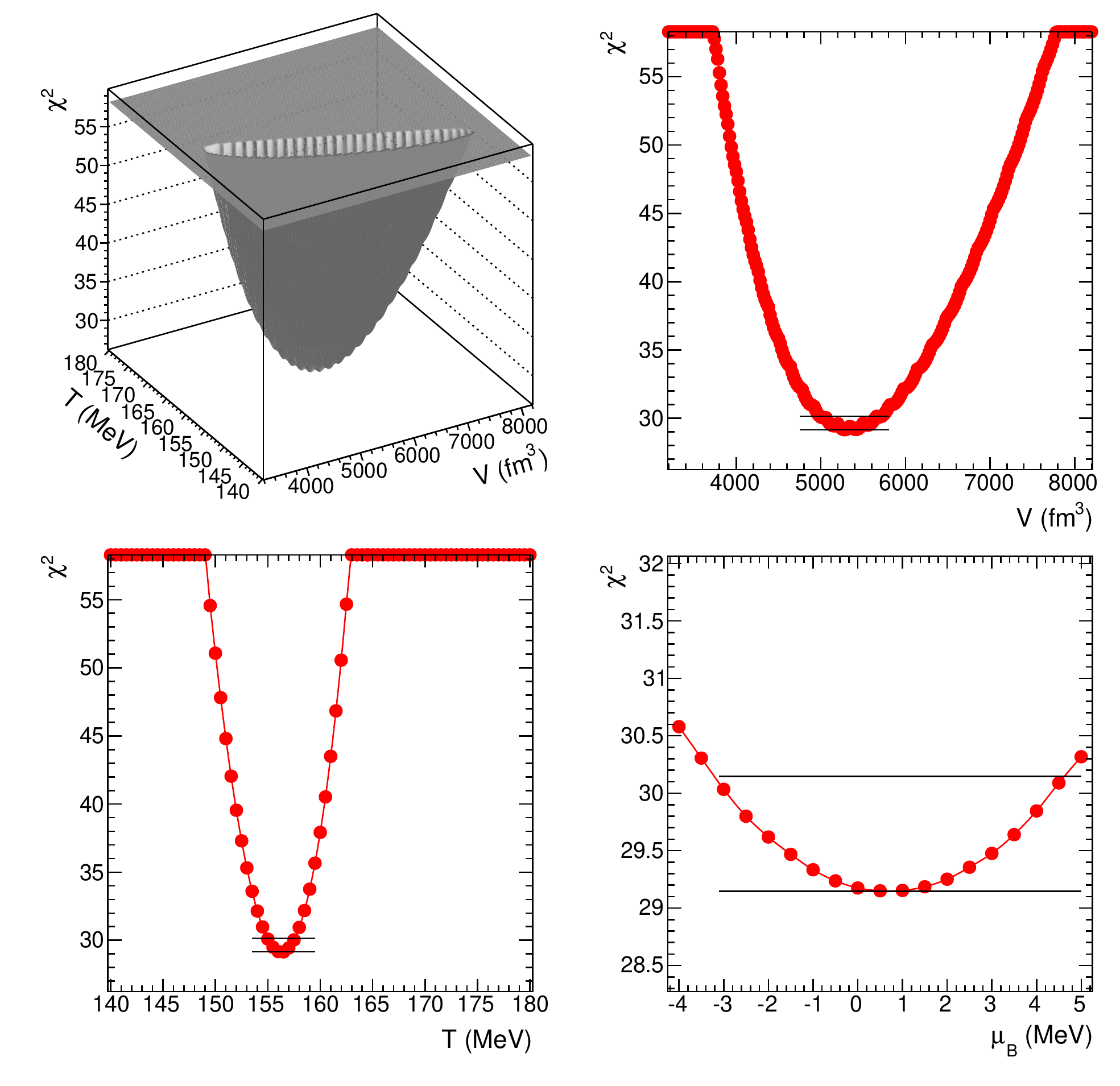}}
\caption{The variation of $\chi^2$ as a function of the fit parameters. The upper-left panel shows $\chi^2$ in the $T-V$ correlation, the other panels the variation of $\chi^2$ for each parameter (the horizontal lines indicate the $\chi^2_{min}$ and $\chi^2_{min}+1$ values). \label{f:chi}}
\end{figure}

We show below, employing our ``standard'' model
\cite{Andronic:2008gu,Stachel:2013zma}, the current
description of the LHC data in central (0-10\%) Pb--Pb collisions at
$\sqrt{s_{_{\rm NN}}}=2.76$ TeV. 
This model is based on the statistical operator for the hadron
resonance gas which was shown earlier  \cite{Andronic:2012ut} to lead to an equation of state in good agreement with lattice QCD calculations.
The best fit values and their uncertainties are: $T=156.5\pm 1.5$ MeV, $\mu_B=0.7\pm 3.8$ MeV, $V_{\Delta y=1}=5280\pm 410$ fm$^3$,
achieved for $\chi^2_{min}=29.1$ per 18 d.o.f., indicating a very good
description of data, which extends over 9 orders of magnitude in hadron
yields, see Fig.~\ref{f:fit} and  Fig.~\ref{f:chi}.
Included in the calculations are the contribution in the yields of $\pi$, $K^\pm$, and $K^0$ from the decays of charmed hadrons, amounting to a relative contribution for the best fit of 0.7\%, 2.9\%, and 3.1\%, respectively.
While we used $R_{meson}=R_{baryon}=0.3$ fm, point-like hadrons lead to a fit with 
same $T$ and $\mu_B$ values, but a volume smaller by about 25\%. 

\begin{figure}[htb]
\begin{tabular}{lr} \begin{minipage}{.5\textwidth}
\includegraphics[width=.9\textwidth]{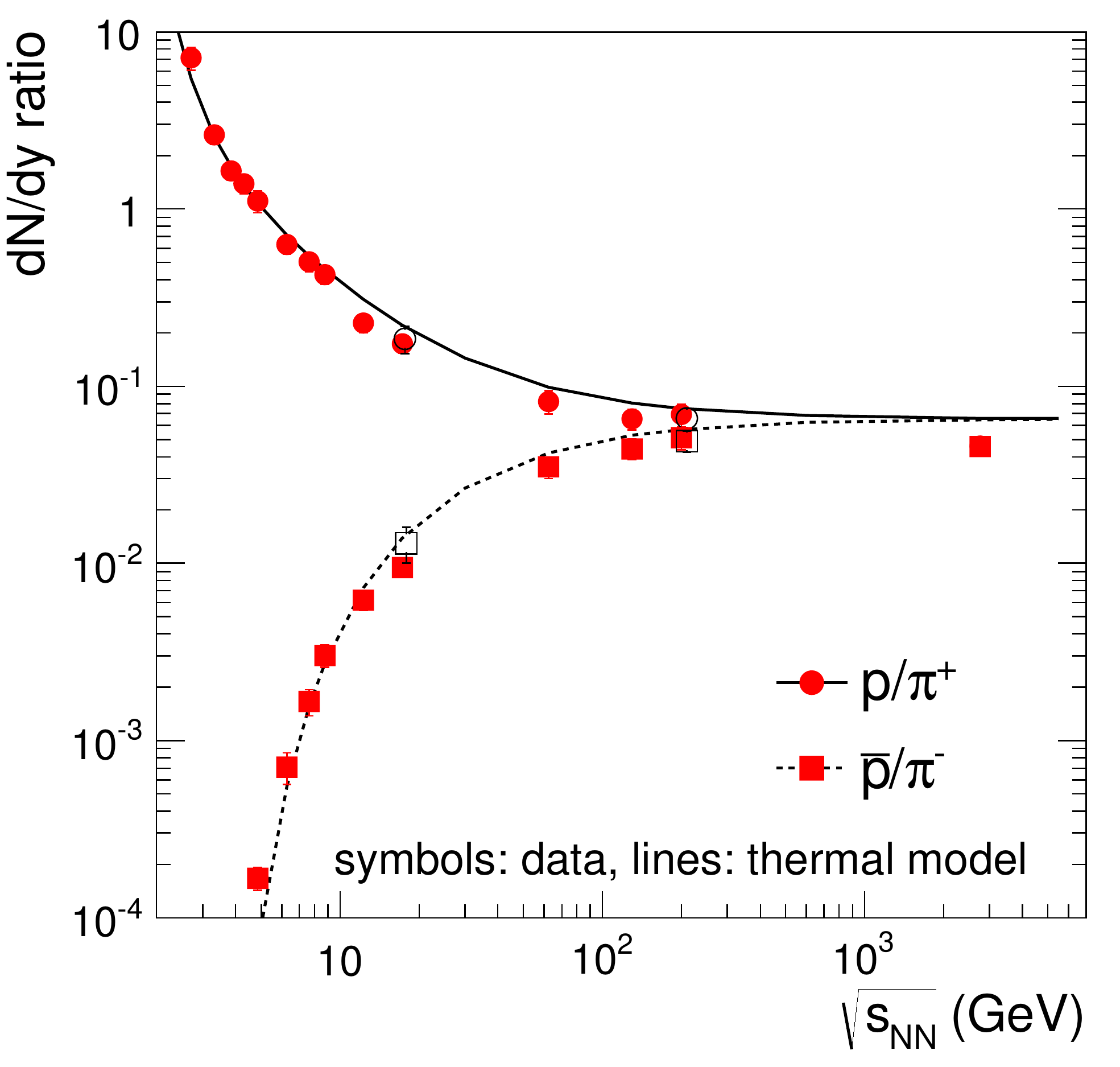}
\end{minipage} & \begin{minipage}{.5\textwidth}
\includegraphics[width=.9\textwidth]{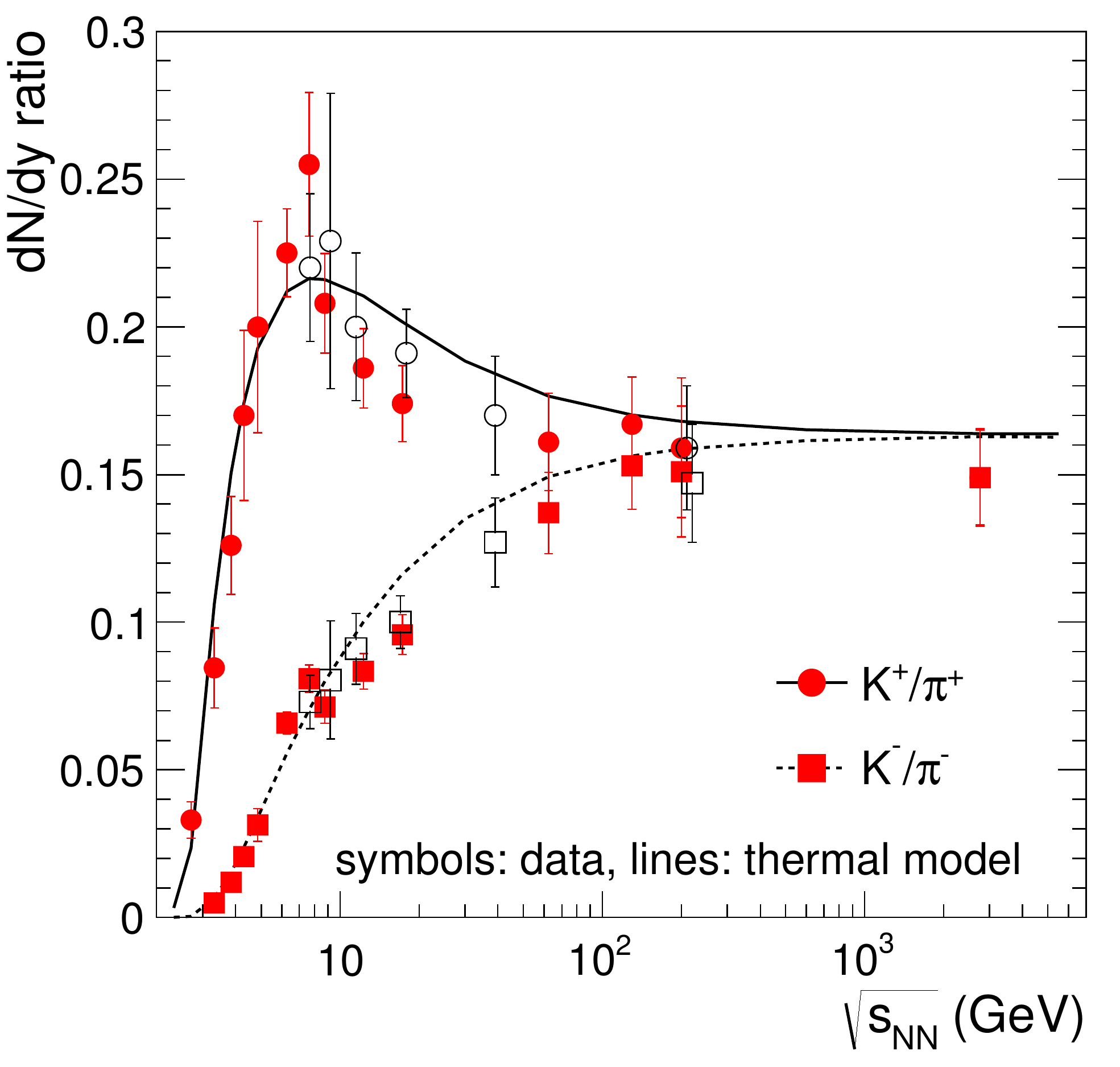}
\end{minipage}\end{tabular}
\caption{
Collision energy dependence of ratios of yields of protons and antiprotons 
(left panel) and of kaons (right panel) to yields of pions.
The symbols are data, the lines are thermal model calculations for energy-dependent parametrizations of $T$ and $\mu_B$ (as in Ref. \cite{Andronic:2008gu}, but with $T_{lim}=159$ MeV).
Full symbols represent the data of NA49 and STAR ($p$, $\bar{p}$ from weak decays subtracted based on the thermal model); at 17 GeV open symbols represent the NA44 data, at 200 GeV BRAHMS data, and for lower energies STAR BES data (preliminary).
\label{f:ratios}
}
\end{figure}

In Fig.~\ref{f:ratios} an illustration is shown of the success of the thermal model in reproducing over a broad energy range the production of (anti)protons and kaons relative to pions.

\begin{figure}[htb]
\vspace{-.2cm}
\begin{minipage}{0.6\textwidth}
\hspace{-.1cm}{\includegraphics[width=.95\textwidth]{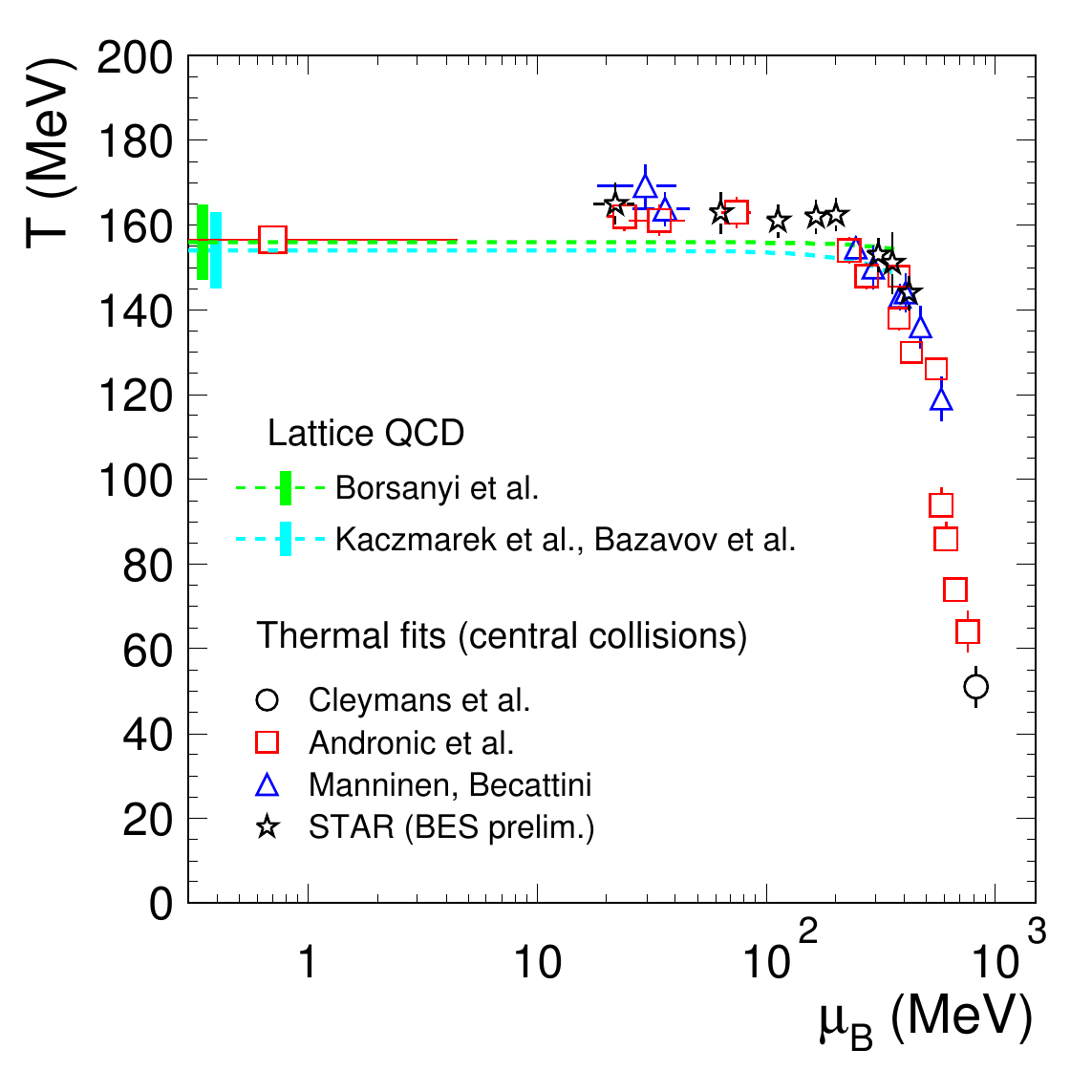}}
\end{minipage}
\begin{minipage}{0.38\textwidth}
\caption{The phase diagram of strongly interacting matter with the points representing the thermal fits of hadron yields at various collision energies \cite{Cleymans:1998yb,Andronic:2008gu,Manninen:2008mg,Abelev:2009bw,Aggarwal:2010pj}.
The crossover $T$ values from lattice QCD \cite{Borsanyi:2010bp,Bazavov:2014noa}
are shown as bands at small $\mu_B$ values. 
The dashed lines represent lattice QCD calculations of the curvature parameter \cite{Kaczmarek:2011zz,Borsanyi:2012cr}.
\label{f:tmu} 
}
\end{minipage} 
\end{figure}

The value of the \mbox{(pseudo-)}critical temperature, $T_c$, at vanishing 
baryochemical potential ($\mu_B$) is currently calculated in lattice 
QCD \cite{Borsanyi:2010bp,Bazavov:2014noa} to be 155$\pm$9 MeV. 
The phenomenological QCD phase diagram is shown in Fig.~\ref{f:tmu}.
Remarkably, at low $\mu_B$ chemical freeze-out coincides with $T_c$,
indicating hadron formation from deconfined matter.
Each point corresponds to a fit of hadron yields in central Au--Au or Pb--Pb 
collisions at a given collision energy. The agreement between the
results from several independent analyses
\cite{Andronic:2008gu,Manninen:2008mg,Abelev:2009bw,Aggarwal:2010pj}
is very good. 
Note that in some cases \cite{Manninen:2008mg,Abelev:2009bw,Aggarwal:2010pj} an
additional fit parameter, the strangeness suppression factor $\gamma_s$, is used
to test possible departure from equilibrium of hadrons containing strange quark(s). Values of $\gamma_s$ (slightly) below unity are found. 
The non-equilibrium model \cite{Letessier:2005qe,Petran:2013qla} leads
to rather different results (smaller $T$ values for small $\mu_B$),
while the model with an extended hadronic phase
\cite{Becattini:2016xct} implies larger $T$ values (neither of these 2
models are shown here).

\section{Charmonium}

The statistical hadronization model outlined above can be applied as
well to production of hadrons with heavy quarks (HQ), charm and bottom
\cite{BraunMunzinger:2000px,Andronic:2006ky}. Given that the HQ
themselves are produced predominantly in primary hard collisions (for charm, $t_{c\bar{c}}\sim 1/2m_c \simeq$ 0.1 fm/$c$), the model describes, in a more explicit way than for lighter quarks, the hadronization stage.
One additional input parameter, the HQ production cross section, is employed, as well as the assumption the HQ survive and thermalize in QGP (thermal, but not chemical equilibrium).
We discuss here the current status of the model description of the J/$\psi$ production at the LHC, quantified via the nuclear modification factor $R_{AA}^{J/\psi}$.

\begin{figure}[htb]
\vspace{-.3cm}
\begin{tabular}{cc} \begin{minipage}{.5\textwidth}
\centerline{\includegraphics[width=.8\textwidth, height=.71\textwidth]{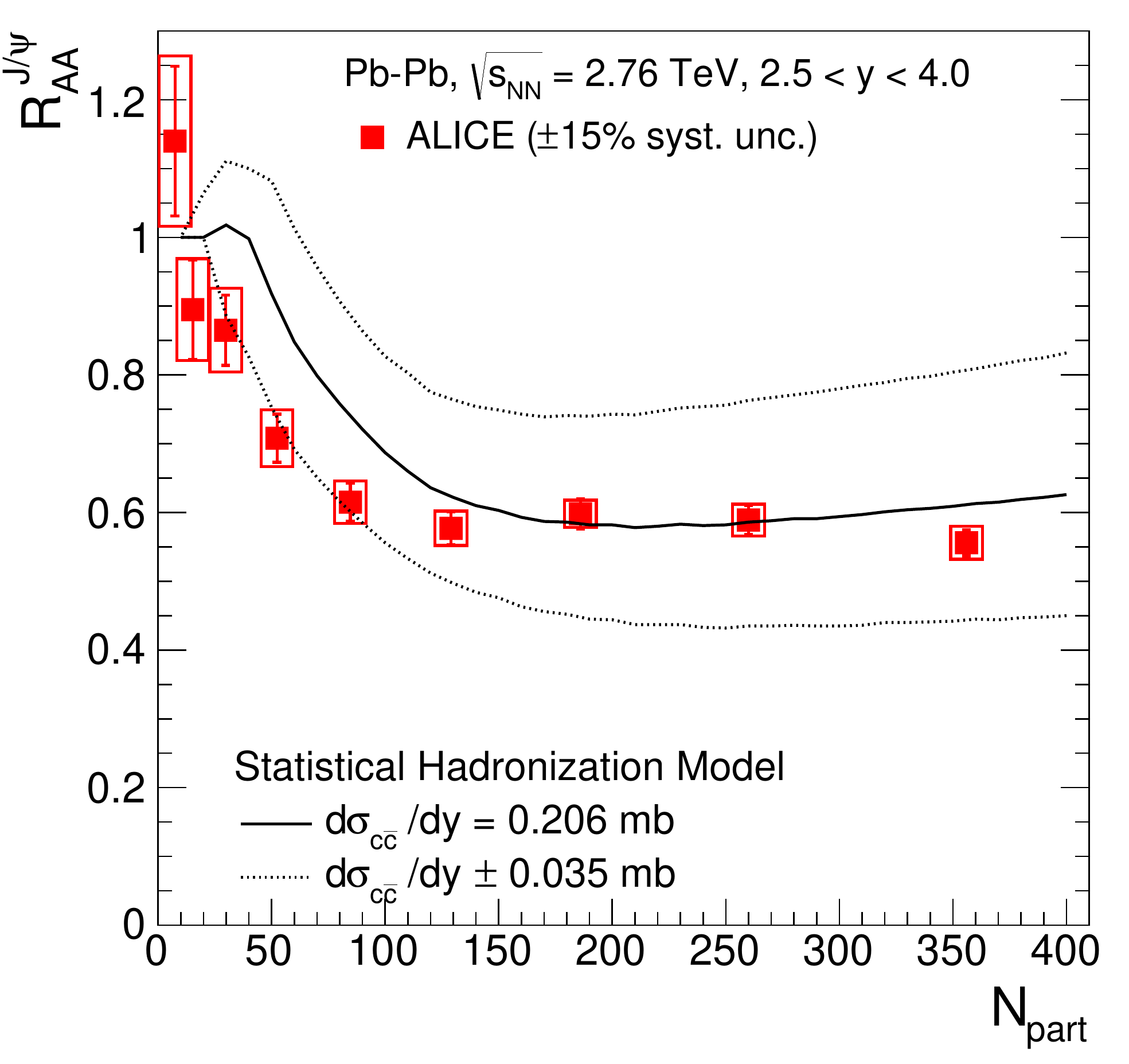}}
\end{minipage} & \begin{minipage}{.5\textwidth}
{\includegraphics[width=.8\textwidth, height=.71\textwidth]{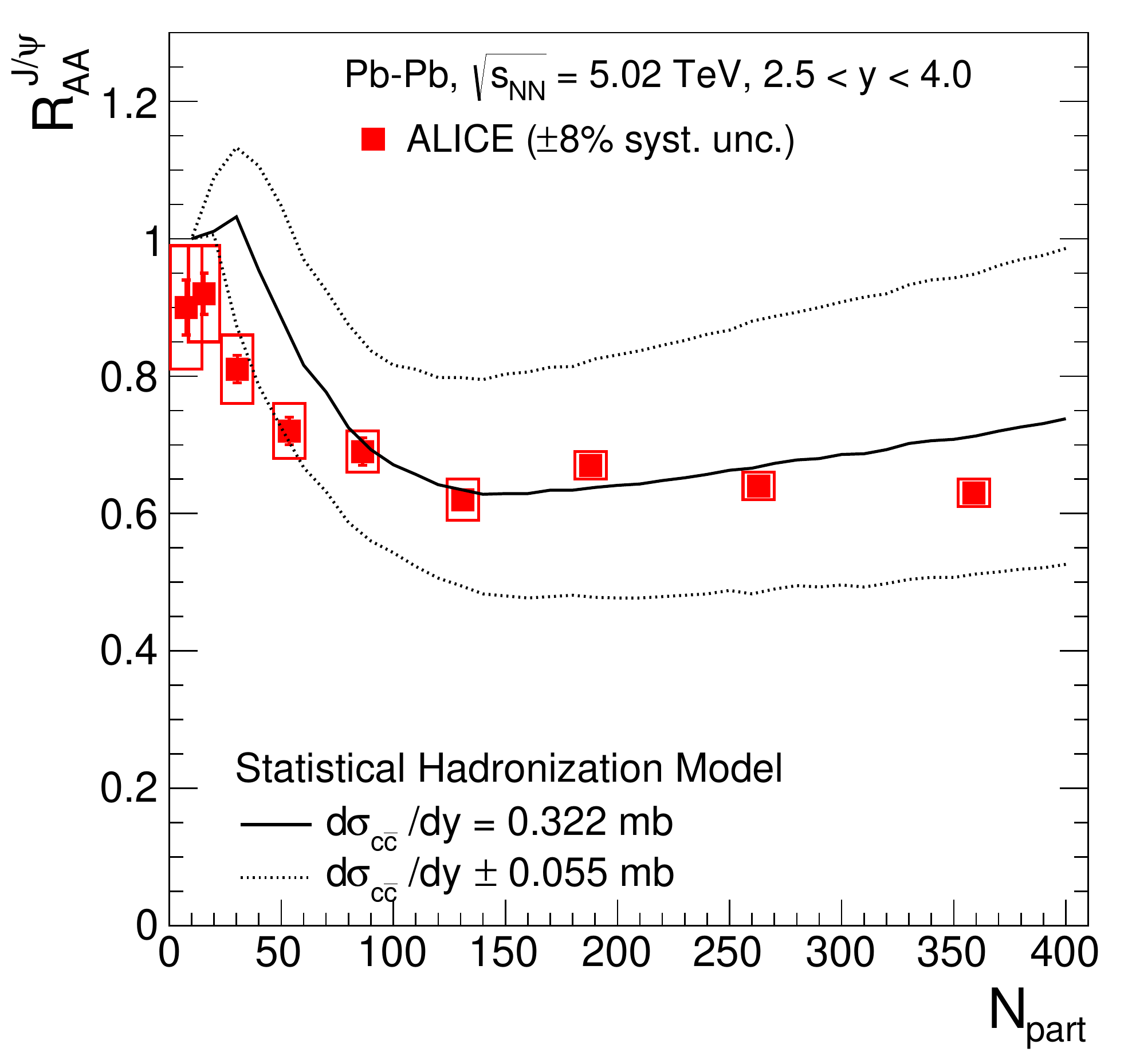}}
\end{minipage}  \\
\begin{minipage}{.5\textwidth}
\centerline{\includegraphics[width=.8\textwidth, height=.71\textwidth]{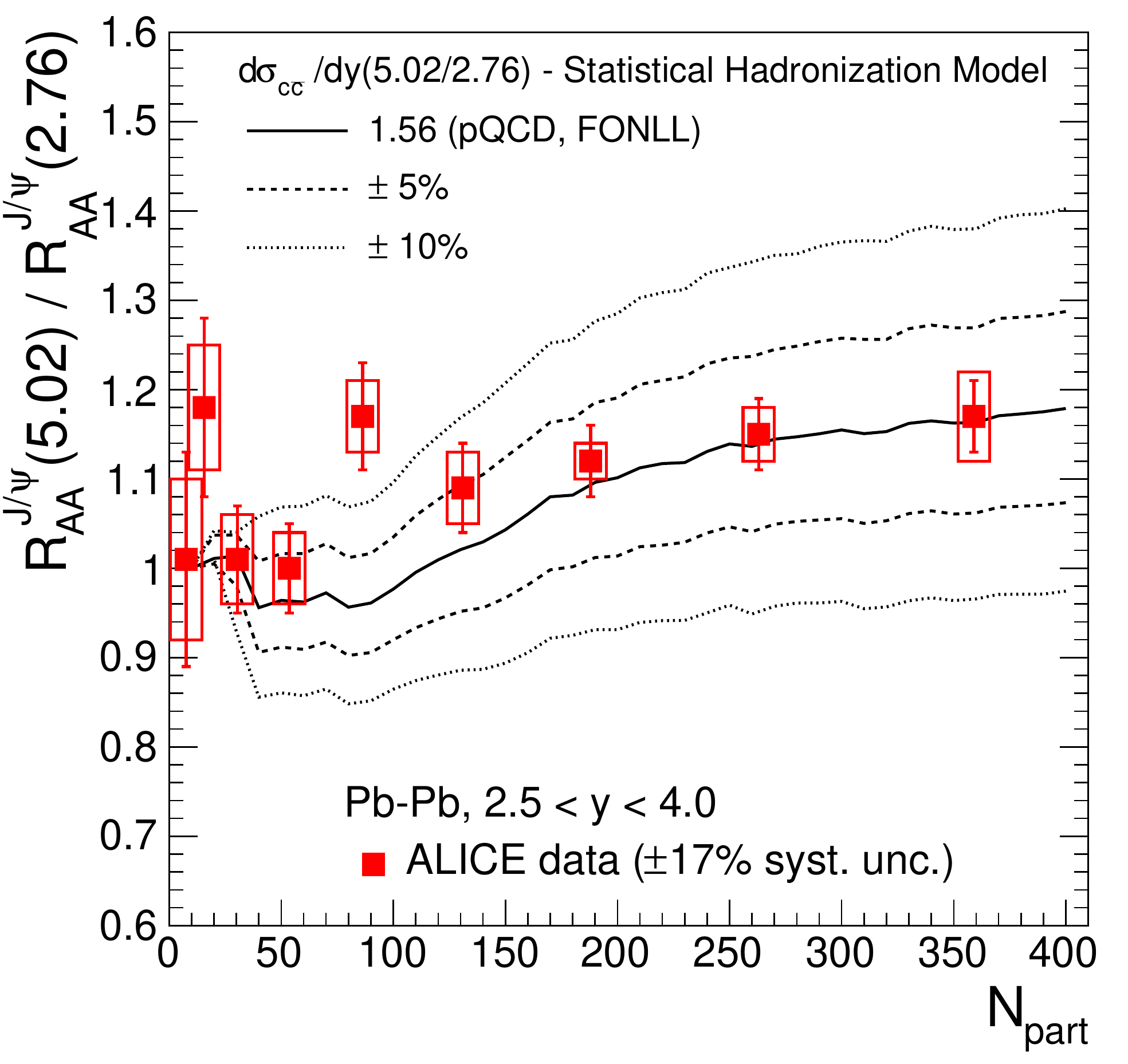}}
\end{minipage} & \begin{minipage}{.5\textwidth}
{\includegraphics[width=.8\textwidth, height=.71\textwidth]{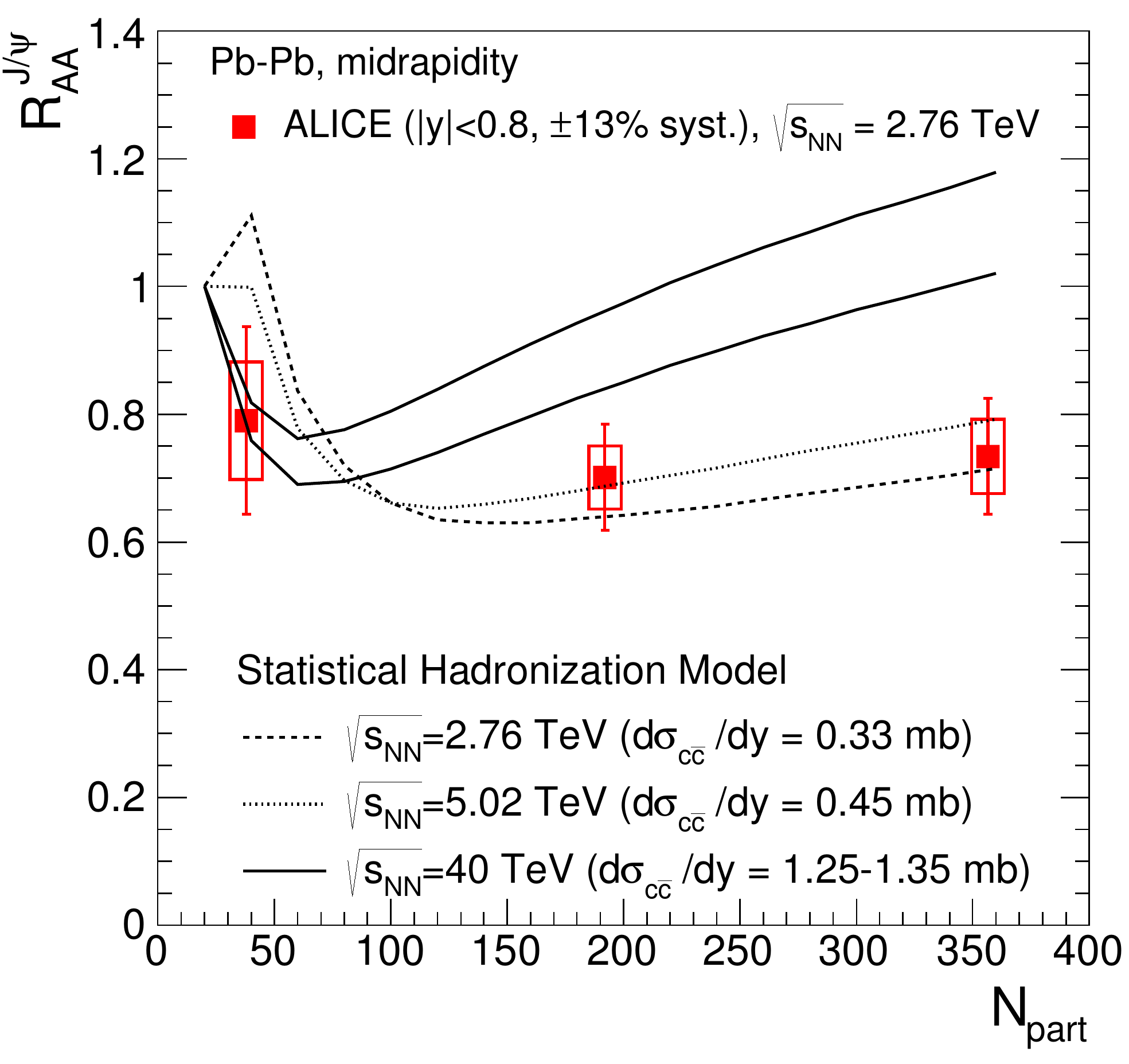}}
\end{minipage} 
\end{tabular} 
\caption{\label{f:jpsi} The centrality dependence of the nuclear modification factor of J/$\psi$ production at the LHC. The model is compared to ALICE data \cite{Abelev:2013ila,Adam:2016rdg} at forward rapidity ($2.5 < y < 4$, upper panel for $\sqrt{s_{_{\rm NN}}}=2.76$ and 5.02 TeV and lower-left panel for the ratio of the 2 energies) and at midrapidity (lower-right panel, for $\sqrt{s_{_{\rm NN}}}=2.76$ TeV and with predictions for 40 TeV \cite{Dainese:2016gch}).}
\end{figure}

Within the statistical hadronization approach a new charmonium
production regime was predicted \cite{BraunMunzinger:2000px,Andronic:2006ky}
for LHC energies. Consequently, the measurement 
was expected to be decisive in clarifying the suppression via the
Debye screening mechanism \cite{Matsui:1986dk} and answering if
(re)generation scenarios are viable production mechanisms. 
The data \cite{Abelev:2013ila,Adam:2016rdg} showed that statistical
generation at the chemical freeze-out is likely the mechanism of J/$\psi$ 
production, as demonstrated by the good agreement of model predictions and 
data, Fig.~\ref{f:jpsi}. The uncertainty in the model predictions is due to 
the $c\bar{c}$ production cross section, obtained by extrapolating the LHCb 
measurement \cite{Aaij:2013mga} in pp collisions at $\sqrt{s}$=7 TeV using 
FONLL pQCD calculations \cite{Cacciari:2012ny} and a shadowing factor from 
EPS09 calculations \cite{Vogt:2015uba}
(0.71$\pm$0.10 for $2.5 < y < 4.0$). Transport models
\cite{Zhao:2011cv,Zhou:2014kka} describe the data as well, assuming
production during the whole lifetime of QGP. However, the tantalizing
possibility of having the J/$\psi$ meson as a probe of the QCD phase
boundary is very appealing.


\vspace{0.2cm}
\textbf{Acknowledgments} This work is part of and supported by the DFG Collaborative Research Centre "SFB 1225 (ISOQUANT)". K.R. acknowledges support by the Polish Science Center (NCN), under Maestro grant DEC-2013/10/A/ST2/00106.

\section*{References}
\bibliography{sqm2016proc}

\providecommand{\newblock}{}
\begin{thebibliography}{10}
\expandafter\ifx\csname url\endcsname\relax
  \def\url#1{{\tt #1}}\fi
\expandafter\ifx\csname urlprefix\endcsname\relax\def\urlprefix{URL }\fi
\providecommand{\eprint}[2][]{\url{#2}}

\bibitem{BraunMunzinger:2008tz}
Braun-Munzinger P and Wambach J 2009 {\em Rev. Mod. Phys.\/} {\bf 81}
  1031--1050 (\textit{Preprint} \eprint{0801.4256})

\bibitem{Cabibbo:1975ig}
Cabibbo N and Parisi G 1975 {\em Phys. Lett. B\/} {\bf 59} 67--69

\bibitem{Hagedorn:1984hz}
Hagedorn R 1985 {\em Lect. Notes Phys.\/} {\bf 221} 53--76

\bibitem{Bazavov:2014xya}
Bazavov A, Ding H~T, Hegde P, Kaczmarek O, Karsch F {\em et~al.\/} 2014 {\em
  Phys. Rev. Lett.\/} {\bf 113} 072001 (\textit{Preprint} \eprint{1404.6511})

\bibitem{Andronic:2005yp}
Andronic A, Braun-Munzinger P and Stachel J 2006 {\em Nucl. Phys. A\/} {\bf
  772} 167--199 (\textit{Preprint} \eprint{nucl-th/0511071})

\bibitem{Becattini:2005xt}
Becattini F, Manninen J and Gazdzicki M 2006 {\em Phys. Rev. C\/} {\bf 73}
  044905 (\textit{Preprint} \eprint{hep-ph/0511092})

\bibitem{Letessier:2005qe}
Letessier J and Rafelski J 2008 {\em Eur. Phys. J. A\/} {\bf 35} 221--242
  (\textit{Preprint} \eprint{nucl-th/0504028})

\bibitem{Petran:2013qla}
Petran M and Rafelski J 2013 {\em Phys. Rev. C\/} {\bf 88} 021901
  (\textit{Preprint} \eprint{1303.0913})

\bibitem{Bazavov:2013dta}
Bazavov A, Ding H~T, Hegde P, Kaczmarek O, Karsch F {\em et~al.\/} 2013 {\em
  Phys. Rev. Lett.\/} {\bf 111} 082301 (\textit{Preprint} \eprint{1304.7220})

\bibitem{Bellwied:2013cta}
Bellwied R, Borsanyi S, Fodor Z, Katz S~D and Ratti C 2013 {\em Phys. Rev.
  Lett.\/} {\bf 111} 202302 (\textit{Preprint} \eprint{1305.6297})

\bibitem{Chatterjee:2013yga}
Chatterjee S, Godbole R and Gupta S 2013 {\em Phys. Lett. B\/} {\bf 727}
  554--557 (\textit{Preprint} \eprint{1306.2006})

\bibitem{Becattini:2016xct}
Becattini F, Steinheimer J, Stock R and Bleicher M 2016  (\textit{Preprint}
  \eprint{1605.09694})

\bibitem{Vovchenko:2015cbk}
Vovchenko V and St{\"o}cker H 2015  (\textit{Preprint} \eprint{1512.08046})

\bibitem{Cleymans:1998yb}
Cleymans J, Oeschler H and Redlich K 1999 {\em Phys. Rev. C\/} {\bf 59} 1663
  (\textit{Preprint} \eprint{nucl-th/9809027})

\bibitem{Agakishiev:2015bwu}
Agakishiev G {\em et~al.\/} (HADES Collaboration) 2016 {\em Eur. Phys. J. A\/}
  {\bf 52} 178 (\textit{Preprint} \eprint{1512.07070})

\bibitem{Vovchenko:2015idt}
Vovchenko V, Begun V~V and Gorenstein M~I 2016 {\em Phys. Rev. C\/} {\bf 93}
  064906 (\textit{Preprint} \eprint{1512.08025})

\bibitem{Das:2016muc}
Das S, Mishra D, Chatterjee S and Mohanty B 2016  (\textit{Preprint}
  \eprint{1605.07748})

\bibitem{Cleymans:2016qnc}
Cleymans J, Hippolyte B, Oeschler H, Redlich K and Sharma N 2016
  (\textit{Preprint} \eprint{1603.09553})

\bibitem{Adam:2015vsf}
Adam J {\em et~al.\/} (ALICE Collaboration) 2016 {\em Phys. Lett. B\/} {\bf
  758} 389--401 (\textit{Preprint} \eprint{1512.07227})

\bibitem{Adam:2016emw}
Adam J {\em et~al.\/} (ALICE Collaboration) 2016  (\textit{Preprint}
  \eprint{1606.07424})

\bibitem{Abelev:2012wca}
Abelev B {\em et~al.\/} (ALICE Collaboration) 2012 {\em Phys. Rev. Lett.\/}
  {\bf 109} 252301 (\textit{Preprint} \eprint{1208.1974})

\bibitem{Abelev:2013vea}
Abelev B {\em et~al.\/} (ALICE Collaboration) 2013 {\em Phys. Rev. C\/} {\bf
  88} 044910 (\textit{Preprint} \eprint{1303.0737})

\bibitem{Abelev:2013xaa}
Abelev B~B {\em et~al.\/} (ALICE Collaboration) 2013 {\em Phys. Rev. Lett.\/}
  {\bf 111} 222301 (\textit{Preprint} \eprint{1307.5530})

\bibitem{ABELEV:2013zaa}
Abelev B~B {\em et~al.\/} (ALICE Collaboration) 2014 {\em Phys. Lett. B\/} {\bf
  728} 216--227 (\textit{Preprint} \eprint{1307.5543})

\bibitem{Abelev:2014uua}
Abelev B~B {\em et~al.\/} (ALICE Collaboration) 2015 {\em Phys. Rev. C\/} {\bf
  91} 024609 (\textit{Preprint} \eprint{1404.0495})

\bibitem{Adam:2015yta}
Adam J {\em et~al.\/} (ALICE Collaboration) 2016 {\em Phys. Lett. B\/} {\bf
  754} 360--372 (\textit{Preprint} \eprint{1506.08453})

\bibitem{Adam:2015vda}
Adam J {\em et~al.\/} (ALICE Collaboration) 2016 {\em Phys. Rev. C\/} {\bf 93}
  024917 (\textit{Preprint} \eprint{1506.08951})

\bibitem{Schuchmann:2102194}
Schuchmann S and Appelshaeuser H 2015 {\em {Modification of $K^{0}_{s}$ and
  $\Lambda(\bar{\Lambda})$ transverse momentum spectra in Pb-Pb collisions at
  $\sqrt{s_{NN}}$ = 2.76 TeV with ALICE}\/} Ph.D. thesis Frankfurt U.

\bibitem{Andronic:2008gu}
Andronic A, Braun-Munzinger P and Stachel J 2009 {\em Phys. Lett. B\/} {\bf
  673} 142--145 [Erratum: Phys. Lett. B 678, 516 (2009)] (\textit{Preprint}
  \eprint{0812.1186})

\bibitem{Stachel:2013zma}
Stachel J, Andronic A, Braun-Munzinger P and Redlich K 2014 {\em J. Phys. Conf.
  Ser.\/} {\bf 509} 012019 (\textit{Preprint} \eprint{1311.4662})

\bibitem{Andronic:2012ut}
Andronic A, Braun-Munzinger P, Stachel J and Winn M 2012 {\em Phys. Lett. B\/}
  {\bf 718} 80--85 (\textit{Preprint} \eprint{1201.0693})

\bibitem{Manninen:2008mg}
Manninen J and Becattini F 2008 {\em Phys. Rev. C\/} {\bf 78} 054901
  (\textit{Preprint} \eprint{0806.4100})

\bibitem{Abelev:2009bw}
Abelev B {\em et~al.\/} (STAR Collaboration) 2010 {\em Phys. Rev. C\/} {\bf 81}
  024911 (\textit{Preprint} \eprint{0909.4131})

\bibitem{Aggarwal:2010pj}
Aggarwal M {\em et~al.\/} (STAR Collaboration) 2011 {\em Phys. Rev. C\/} {\bf
  83} 034910 (\textit{Preprint} \eprint{1008.3133})

\bibitem{Borsanyi:2010bp}
Borsanyi S {\em et~al.\/} (Wuppertal-Budapest Collaboration) 2010 {\em JHEP\/}
  {\bf 1009} 073 (\textit{Preprint} \eprint{1005.3508})

\bibitem{Bazavov:2014noa}
Bazavov A, Bhattacharya T, DeTar C, Ding H~T, Gottlieb S {\em et~al.\/}
  (HotQCD) 2014 {\em Phys. Rev. D\/} {\bf 90} 094503 (\textit{Preprint}
  \eprint{1407.6387})

\bibitem{Kaczmarek:2011zz}
Kaczmarek O, Karsch F, Laermann E, Miao C, Mukherjee S {\em et~al.\/} 2011 {\em
  Phys. Rev. D\/} {\bf 83} 014504 (\textit{Preprint} \eprint{1011.3130})

\bibitem{Borsanyi:2012cr}
Borsanyi S, Endrodi G, Fodor Z, Katz S, Krieg S {\em et~al.\/} 2012 {\em
  JHEP\/} {\bf 1208} 053 (\textit{Preprint} \eprint{1204.6710})

\bibitem{BraunMunzinger:2000px}
Braun-Munzinger P and Stachel J 2000 {\em Phys. Lett. B\/} {\bf 490} 196--202
  (\textit{Preprint} \eprint{nucl-th/0007059})

\bibitem{Andronic:2006ky}
Andronic A, Braun-Munzinger P, Redlich K and Stachel J 2007 {\em Nucl. Phys.
  A\/} {\bf 789} 334--356 (\textit{Preprint} \eprint{nucl-th/0611023})

\bibitem{Abelev:2013ila}
Abelev B~B {\em et~al.\/} (ALICE Collaboration) 2014 {\em Phys. Lett. B\/} {\bf
  743} 314--327 (\textit{Preprint} \eprint{1311.0214})

\bibitem{Adam:2016rdg}
Adam J {\em et~al.\/} (ALICE Collaboration) 2016  (\textit{Preprint}
  \eprint{1606.08197})

\bibitem{Dainese:2016gch}
Dainese A {\em et~al.\/} 2016  (\textit{Preprint} \eprint{1605.01389})

\bibitem{Matsui:1986dk}
Matsui T and Satz H 1986 {\em Phys. Lett. B\/} {\bf 178} 416

\bibitem{Aaij:2013mga}
Aaij R {\em et~al.\/} (LHCb Collaboration) 2013 {\em Nucl. Phys. B\/} {\bf 871}
  1--20 (\textit{Preprint} \eprint{1302.2864})

\bibitem{Cacciari:2012ny}
Cacciari M, Frixione S, Houdeau N, Mangano M~L, Nason P {\em et~al.\/} 2012
  {\em JHEP\/} {\bf 1210} 137 (\textit{Preprint} \eprint{1205.6344})

\bibitem{Vogt:2015uba}
Vogt R 2015 {\em Phys. Rev. C\/} {\bf 92} 034909 (\textit{Preprint}
  \eprint{1507.04418})

\bibitem{Zhao:2011cv}
Zhao X and Rapp R 2011 {\em Nucl. Phys. A\/} {\bf 859} 114--125
  (\textit{Preprint} \eprint{1102.2194})

\bibitem{Zhou:2014kka}
Zhou K, Xu N, Xu Z and Zhuang P 2014 {\em Phys. Rev. C\/} {\bf 89} 054911
  (\textit{Preprint} \eprint{1401.5845})

\end{thebibliography}

\end{document}